\documentclass{article}
\usepackage{amssymb}
\usepackage{array}
\usepackage{amsfonts}
\usepackage{amsmath}
\usepackage{rotate,epsfig}
\usepackage{amsfonts}
\usepackage{lscape}
\usepackage{graphicx}
\usepackage{xcolor}

\begin{document}

\title{Using a rank-based design in estimating prevalence of breast cancer}
\author{ {M. Mahdizadeh}${}^{\rm a,}$\footnote{Corresponding author.\newline {\it E-mail addresses:}
\texttt{mahdizadeh.m@hsu.ac.ir; mahdizadeh.m@live.com} (M. Mahdizadeh), \texttt{e.zamanzade@sci.ui.ac.ir; ehsanzamanzadeh@yahoo.com} (Ehsan Zamanzade)} \,and { Ehsan Zamanzade}${}^{\rm b}$ {\vspace{2mm}}\\
{\normalsize {\it $~^{\rm a}$Department of Statistics, Hakim Sabzevari University,  P.O. Box 397, Sabzevar, Iran}}\vspace{-0.1cm}\\
{\normalsize {\it $~^{\rm b}$Department of Statistics, Faculty of Mathematics and Statistics,}}\vspace{-0.1cm}\\
{\normalsize {\it  University of Isfahan, Isfahan 81746-73441, Iran}}}\vspace{-0.1cm}
\maketitle

\begin{abstract}
\noindent It is highly important for governments and health organizations to monitor the prevalence of breast cancer as a leading source of cancer-related death among women. However, the accurate diagnosis of this disease is expensive, especially in developing countries. This article concerns a cost-efficient method for estimating prevalence of breast cancer, when diagnosis is based on a comprehensive biopsy procedure. Multistage ranked set sampling (MSRSS) is utilized to develop a proportion estimator. This design employs some visually assessed cytological covariates, which are pertinent to determination of breast cancer, so as to provide the experimenter with a more informative sample. Theoretical properties of the proposed estimator are explored. Evidence from numerical studies is reported. The developed procedure can be substantially more efficient than its competitor in simple random sampling (SRS). In some situations, the proportion estimation in MSRSS needs around 76\% fewer observations than that in SRS, given a precision level. Thus, using MSRSS may lead to a considerable reduction in cost with respect to SRS. In many medical studies, e.g. diagnosing breast cancer based on a full biopsy procedure, exact quantification is difficult (costly and/or time-consuming), but the potential sample units can be ranked fairly accurately without actual measurements. In this setup, multistage ranked set sampling is an appropriate design for developing cost-efficient statistical methods.

\end{abstract}

\section{Introduction}
Breast cancer is the most commonly occurring female cancer, and a leading source of cancer-related deaths among women worldwide. It is thought to be a disease of the developed world, while it is prevalent in developing countries as well. Breast cancer survival rate in developed countries is higher than that in middle/low-income countries, where the lack of early detection programs with good coverage is pronounced. In developing countries, the situation is deteriorated by the lack of adequate diagnosis and treatment facilities. Many studies have been conducted about different aspects of breast cancer (Shapiro, 2018).

Although the causes of breast cancer are not fully understood, researchers concur that certain factors raise a person's risk of developing this disease. These factors include age, family and personal health history, genetics, and hormonal factor, among others. Adopting proper prevention measures, at different levels, may reduce incidence rate of the disease in the long term.

Unfortunately, there is no definitive method of preventing breast cancer. The early detection, however, is a crucial step for the successful management of the disease. It allows for different treatment options, thereby increasing survival chance, and improving quality of life. In addition, this may cut the treatment costs which are high for both the involved person and for society as a whole.

Some standard tests which can be utilized to diagnose breast cancer include breast exam, mammogram, breast ultrasound, and biopsy. The last one is the only definite way to diagnose breast cancer. In this method, tissues or cells are removed from the body in order to be tested in a lab. Then, a pathologist verifies whether the sample contains cancer cells. A comprehensive biopsy procedure is costly, and processing the results takes time. Fine needle aspiration (FNA) biopsy is the least expensive technique of tissue sampling, which provides reliable results. This belongs to a group of less invasive methods for sampling a breast lesion to determine if it is cancerous or benign. The FNA biopsy has been successfully applied in early detection of breast cancer. Owing to this diagnostic tool, many unjustified major surgical (open) biopsies are avoided.

Suppose we are interested in estimating the prevalence of breast cancer in a given population, using a full biopsy procedure. Taking cost considerations into account, it is interesting to employ a sampling design that enables us to draw inference about the population proportion based on a possibly small sample. Toward this end, rank-based sampling methods can be efficiently adopted. They are applicable in settings where exact quantification is difficult (costly and/or time-consuming), while informal ranking of the potential sample units can be done fairly accurately and easily. The rankings are performed by judgment or through use of an easily available covariate, and they need not be totally free of errors. For example, in the context breast cancer, the FNA biopsy is a quick and a relatively cheap test which yields some visually assessed cytological characteristics (covariates). These covariates can be used to rank the patients according to the probability of having cancerous tumors in a rank-based sampling method. The informal ranking process assists the experimenter to focus attention toward the actual measurement of more representative units in the population, thereby enhancing precision of the estimation.

Ranked set sampling (RSS) is a ranked-based design, due to McIntyre (1952, 2005) which is a cost-efficient alternative to simple random sampling (SRS). It has been recently applied in a variety of disciplines, including agriculture (Mahdizadeh and Zamanzade, 2018), auditing (Gemayel et al., 2012), and medicine (Zamanzade and Mahdizadeh, 2017) among others. Bouza-Herrera and Al-Omari (2019) collect some new developments in this area. In this article, we deal with estimating the prevalence of breast cancer using a generalization of RSS, i.e. multistage ranked set sampling (MSRSS).

In Section 2, proportion estimator under MSRSS is presented, and its properties are studied. Section 3 contains results of a simulation study performed to investigate finite sample performance of the proposed estimator. Some illustrations using real data are also included. Final conclusions and directions for future research are provided in Section 4. Proofs are collected in two appendixes.

\section{Methods}
According to RSS scheme, a sample of size $N=nm$, using set size $m$, is obtained as follows:
\begin{enumerate}
\item First, $m^2$ units are identified from the population.
\item Next, the $m^2$ units are randomly divided into $m$ sets of size $m$.
\item The elements of the $i$th ($i=1,\ldots,m$) set are ordered, and the unit with judgment rank $i$ is identified.
\item The $m$ units identified in step 3 are actually measured.
\item Finally, steps 1-4 are repeated for $n$ cycles.
\end{enumerate}
The ranking mechanism in step 3 does not involve actual quantifications of the attribute of interest. This can be based on expert opinion, or covariates' information. The final sample may be denoted as $\{X_{[i]j}: i=1,\ldots,m\,;j=1,\ldots,n \}$, where ${{X}_{[i]j}}$ is the $i$th judgment order statistic in the $j$th cycle. This sample is more informative than a sample of size $N$ collected by SRS, and thus it often improves statistical inference. The better ranking quality, the higher amount of improvement. Perfect ranking is the situation that ranking errors are absent. The ranking scheme is said to be consistent if the same ranking mechanism is applied to all sets of size $m$.

In order to clarify RSS, we describe drawing a ranked set sample using $m=3$ and $n=1$. First, 9 sample units are identified from the population, and randomly divided into 3 sets of size 3. The three sets are denoted by
$$\left\{ U_1^1, U_2^1, U_3^1 \right\}, \quad \left\{ U_1^2, U_2^2, U_3^2 \right\}, \quad \left\{ U_1^3, U_2^3, U_3^3 \right\},$$
where $U_i^j$ ($i,j=1,2,3$) is the $i$th units in the $j$th set. In each set, the units are ordered with respect to the variable of interest. This step results in
$$\left\{ \boxed{U_{[1]}^1}, U_{[2]}^1, U_{[3]}^1 \right\}, \quad \left\{ U_{[1]}^2, \boxed{U_{[2]}^2}, U_{[3]}^2 \right\}, \quad \left\{ U_{[1]}^3, U_{[2]}^3, \boxed{U_{[3]}^3} \right\},$$
where $U_{[i]}^j$ ($i,j=1,2,3$) shows the unit with judgment rank $i$ in the $j$th set. Finally, the ranked set sample is obtained by quantifying the variable of interest for elements of the set $\left\{ U_{[1]}^1, U_{[2]}^2, U_{[3]}^3 \right\}$.

Many RSS-based procedures have been developed for discrete and continuous data; see Wolfe (2012) for a recent review. In the following, we focus on binary data. Terpstra and Liudahl (2004), and Chen et al. (2005) addressed point estimation for the population proportion $p$, while Terpstra and Miller (2006), and Terpstra and Wang (2008) dealt with interval estimation problem. Let ${{X}_{[i]j}}$ be either 0 or 1 representing a failure or success, respectively. Then, proportion estimator in RSS is given by
$$\hat{p}_{\textrm{RSS}}=\frac{1}{n m} \sum_{i=1}^m \sum_{j=1}^n X_{[i]j}.$$
Suppose the proportion estimator based on a simple random sample of size $N$ is denoted by $\hat{p}_{\textrm{SRS}}$. The next result states important properties of this estimator, which holds regardless of the ranking quality. It can be found in the literature, but the proof has not been detailed, as far as we know.\\
{\bf Proposition 1:} {\it Let $\{X_{[i]j}: i=1,\ldots,m\,;j=1,\ldots,n \}$ be a ranked set sample, drawn based on a consistent ranking scheme, from a population with proportion $p$. If $\hat{p}_{\textrm{RSS}}$ is defined as above and $p_{[i]}=E\left( X_{[i]1} \right)$, then\\
a) $E\left( \hat{p}_{\textrm{RSS}} \right)=p$ and $Var\left( \hat{p}_{\textrm{RSS}} \right)=\sum_{i=1}^m p_{[i]} \left( 1-p_{[i]} \right)/(n m^2)$.\\
b) $Var\left( \hat{p}_{\textrm{RSS}} \right) \leq Var\left( \hat{p}_{\textrm{SRS}} \right)$.\\
c) As $n$ tends to infinity,
$$\sqrt{N}\left( \hat{p}_{\textrm{RSS}}-p \right) \stackrel{d}{\rightarrow} \mathcal{N}\left(0, \frac{1}{m}\sum_{i=1}^m p_{[i]} \left( 1-p_{[i]} \right) \right),$$
where $\stackrel{d}{\rightarrow}$ denotes convergence in distribution.}

The original RSS scheme has been tailored to propose more efficient designs in specific situations. MSRSS scheme, due to Al-Saleh and Al-Omari (2002), is an interesting generalization that allows attaining higher efficiency, given a fixed set size. An $r$th stage ranked set sample of size $N=mn$, using set size $m$, is obtained as follows:
\begin{enumerate}
\item First, $m^{r+1}$ units are identified from the population.
\item Next, the $m^{r+1}$ units are randomly divided into $m^{r-1}$ sets of size $m^2$.
\item Steps 1 and 2 of RSS algorithm are done on each set in step 2 to have a (judgement) ranked set of size $m$. This yields $m^{r-1}$ (judgement) ranked sets of size $m$.
\item Step 3 is done on the $m^{r-1}$ ranked sets to have $m^{r-2}$ second stage (judgement) ranked sets of size $m$.
\item Step 3 is repeated until ending in an $r$th stage (judgement) ranked set of size $m$.
\item The $m$ units identified in step 5 are actually measured.
\item Finally, steps 1-6 are repeated for $n$ cycles.
\end{enumerate}
The resulting multistage ranked set sample is denoted by $\{X_{[i]j}^{(r)}: i=1,\ldots,m\,;j=1,\ldots,n \}$, where $X_{[i]j}^{(r)}$ is the $i$th judgment order statistic in the $j$th cycle. Apparently, MSRSS with $r=1$ is simply the basic RSS.

We now illustrate drawing a multistage ranked set sample using $r=2$, $m=3$, and $n=1$. First, 27 sample units are identified from the population, and randomly divided into 3 sets of size 9. The three sets are denoted by
\begin{equation*}
\begin{pmatrix}
V_{11}^1 & V_{12}^1 & V_{13}^1 \\
V_{21}^1 & V_{22}^1 & V_{23}^1 \\
V_{31}^1 & V_{32}^1 & V_{33}^1
\end{pmatrix}, \quad
\begin{pmatrix}
V_{11}^2 & V_{12}^2 & V_{13}^2 \\
V_{21}^2 & V_{22}^2 & V_{23}^2 \\
V_{31}^2 & V_{32}^2 & V_{33}^2
\end{pmatrix}, \quad
\begin{pmatrix}
V_{11}^3 & V_{12}^3 & V_{13}^3 \\
V_{21}^3 & V_{22}^3 & V_{23}^3 \\
V_{31}^3 & V_{32}^3 & V_{33}^3
\end{pmatrix},
\end{equation*}
where $V_{ij}^k$ ($i,j,k=1,2,3$) is the unit in the $i$th row and $j$th column of the $k$th set. In each set, the units of each row are ordered with respect to the variable of interest. This step yields
\begin{equation*}
\begin{pmatrix}
\boxed{V_{[11]}^1} & V_{[12]}^1 & V_{[13]}^1 \\
V_{[21]}^1 & \boxed{V_{[22]}^1} & V_{[23]}^1 \\
V_{[31]}^1 & V_{[32]}^1 & \boxed{V_{[33]}^1}
\end{pmatrix}, \quad
\begin{pmatrix}
\boxed{V_{[11]}^2} & V_{[12]}^2 & V_{[13]}^2 \\
V_{[21]}^2 & \boxed{V_{[22]}^2} & V_{[23]}^2 \\
V_{[31]}^2 & V_{[32]}^2 & \boxed{V_{[33]}^2}
\end{pmatrix}, \quad
\begin{pmatrix}
\boxed{V_{[11]}^3} & V_{[12]}^3 & V_{[13]}^3 \\
V_{[21]}^3 & \boxed{V_{[22]}^3} & V_{[23]}^3 \\
V_{[31]}^3 & V_{[32]}^3 & \boxed{V_{[33]}^3}
\end{pmatrix},
\end{equation*}
where $V_{[ij]}^k$ ($i,j,k=1,2,3$) shows the unit with judgment rank $j$ in the $i$th row of the $k$th set. Next, the units in the sets
$$S_1=\left\{ V_{[11]}^1, V_{[22]}^1, V_{[33]}^1 \right\}, \quad S_2=\left\{ V_{[11]}^2, V_{[22]}^2, V_{[33]}^2 \right\}, \quad S_3=\left\{ V_{[11]}^3, V_{[22]}^3, V_{[33]}^3 \right\},$$
are ordered. Finally, the 2nd stage ranked set sample is obtained by quantifying the variable of interest for elements of the set $\left\{ \mathcal{V}_{[1]}^1, \mathcal{V}_{[2]}^2, \mathcal{V}_{[3]}^3 \right\}$, where $\mathcal{V}_{[i]}^i$ ($i=1,2,3$) is the unit with judgment rank $i$ in $S_i$.

MSRSS has been applied in estimating the population mean by Al-Saleh and Al-Omari (2002). Frey and Feeman (2018), and Mahdizadeh and Zamanzade (2017b, 2019a) are examples of recent works based on this design. To the best of our knowledge, proportion estimation in MSRSS has not been investigated in the literature. In particular, this is a frequently used procedure in medical studies.

We propose proportion estimator in MSRSS as
$$\hat{p}_{\textrm{MSRSS}}^{(r)}=\frac{1}{n m} \sum_{i=1}^m \sum_{j=1}^n X_{[i]j}^{(r)}.$$
Properties of this estimator are derived in analogy with the sample mean in MSRSS. The main results are summarized in the next proposition.\\
{\bf Proposition 2:} {\it Let $\{X_{[i]j}^{(r)}: i=1,\ldots,m\,;j=1,\ldots,n \}$ be a multistage ranked set sample, drawn based on a consistent ranking scheme, from a population with proportion $p$. If $\hat{p}_{\textrm{MSRSS}}^{(r)}$ is defined as above and $p_{[i]}^{(r)}=E\left( X_{[i]1}^{(r)} \right)$, then\\
a) $E\left( \hat{p}_{\textrm{MSRSS}}^{(r)} \right)=p$ and $Var\left( \hat{p}_{\textrm{MSRSS}}^{(r)} \right)=\sum_{i=1}^m p_{[i]}^{(r)} \left( 1-p_{[i]}^{(r)} \right)/(n m^2)$.\\
b) $Var\left( \hat{p}_{\textrm{MSRSS}}^{(r)} \right) \leq Var\left( \hat{p}_{\textrm{SRS}} \right)$.\\
c) The variance of $\hat{p}_{\textrm{MSRSS}}^{(r)}$ is decreasing in $r$, if the perfect ranking is assumed.\\
d) As $n$ tends to infinity,
$$\sqrt{N}\left( \hat{p}_{\textrm{MSRSS}}^{(r)}-p \right) \stackrel{d}{\rightarrow} \mathcal{N}\left(0, \frac{1}{m}\sum_{i=1}^m p_{[i]}^{(r)} \left( 1-p_{[i]}^{(r)} \right) \right),$$
where $\stackrel{d}{\rightarrow}$ denotes convergence in distribution.}

\section{Results}
\subsection{Simulation study}
To investigate finite-sample properties of the proposed estimator, we conducted a Monte Carlo simulation. Since the proportion estimators in SRS and RSS are both unbiased, relative efficiency (RE) of $\hat{p}_{\textrm{MSRSS}}^{(r)}$ with respect to $\hat{p}_{\textrm{SRS}}$ can be defined as
$$\textrm{RE}=\frac{\textrm{Var}\!\left( \hat{p}_{\textrm{SRS}} \right)}{\textrm{Var}\!\left( \hat{p}_{\textrm{MSRSS}}^{(r)} \right)}$$
Comparing the variance of $\hat{p}_{\textrm{MSRSS}}^{(r)}$ in Proposition 2 (a) with that of $\hat{p}_{\textrm{SRS}}$, it is easily seen that the above RE does not depend on the number of cycles. In our comparisons, we thus assumed that $n=1$. Also, $m \in \{3,4,5\}$, $r \in \{1,2,3,4\}$, and $p \in [0,1]$ were selected.

The numerator of the RE has a simple formula, i.e., $Var\left( \hat{p}_{\textrm{SRS}} \right)=p(1-p)/N$. In order to determine the RE, the variance of $\hat{p}_{\textrm{MSRSS}}^{(r)}$ should be approximated, as it is not available in explicit form. To this end, 100,000 samples are drawn in MSRSS, and the proportion estimator is computed from each sample. Finally, sample variance of the resulting values provides an approximation for the variance of $\hat{p}_{\textrm{MSRSS}}^{(r)}$.  In generating multistage ranked set samples, we need a model that allows to consider possibility of the judgment ranking errors. In RSS literature, such models are known as imperfect ranking models; see Frey (2007), for example. In the following, we describe an extension of the imperfect ranking model developed by Dell and Clutter (1972).

\begin{figure}[h!]
\centering
\includegraphics[]{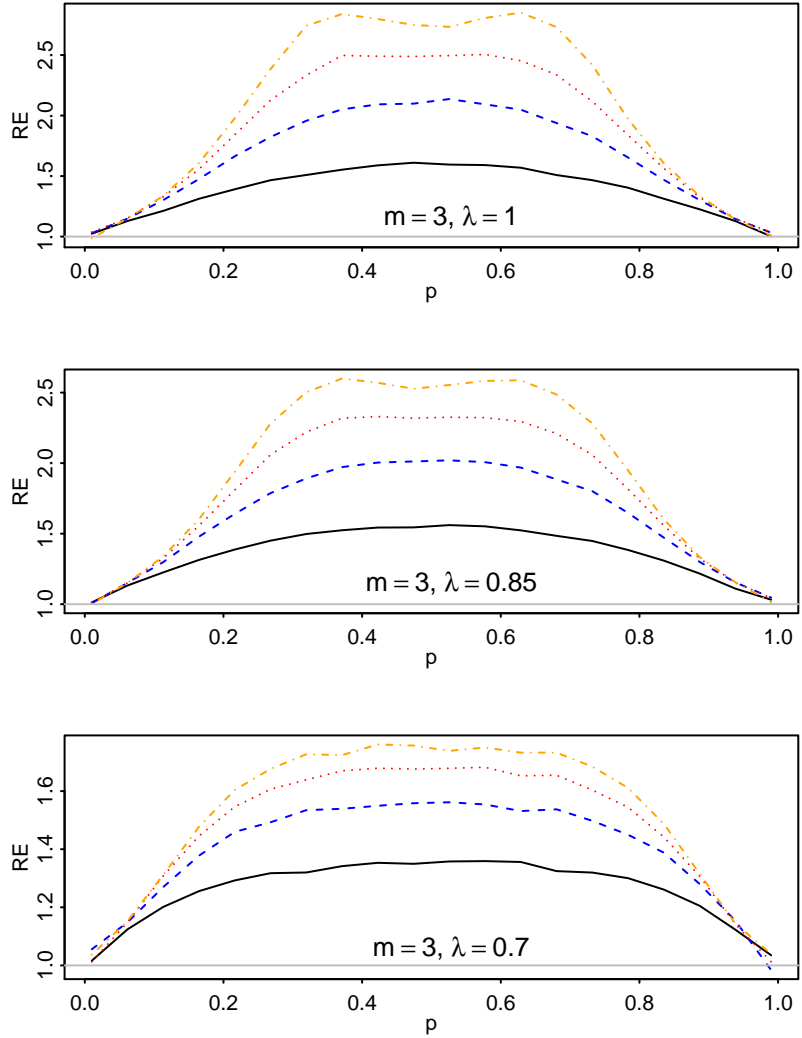}
\caption{The estimated RE for $m=3$ and $\lambda \in \{0.7,0.85,1\}$, where black/solid, blue/dashed, red/dotted, and orange/curves relate to $r=1$,  $r=2$, $r=3$, and $r=4$, respectively.}
\end{figure}

\begin{figure}[h!]
\centering
\includegraphics[]{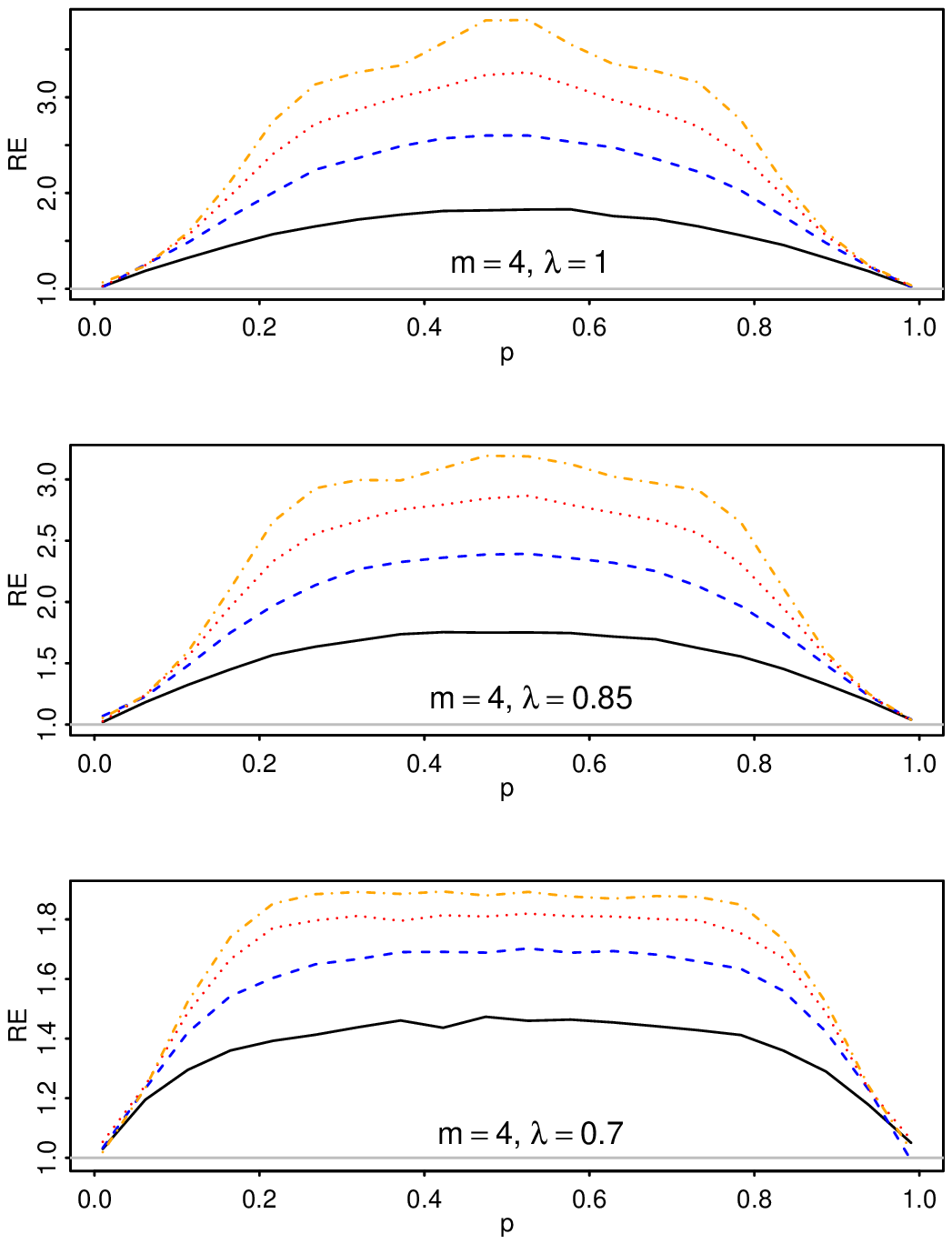}
\caption{The estimated RE for $m=4$ and $\lambda \in \{0.7,0.85,1\}$, where black/solid, blue/dashed, red/dotted, and orange/curves relate to $r=1$,  $r=2$, $r=3$, and $r=4$, respectively.}
\end{figure}

\begin{figure}[h!]
\centering
\includegraphics[]{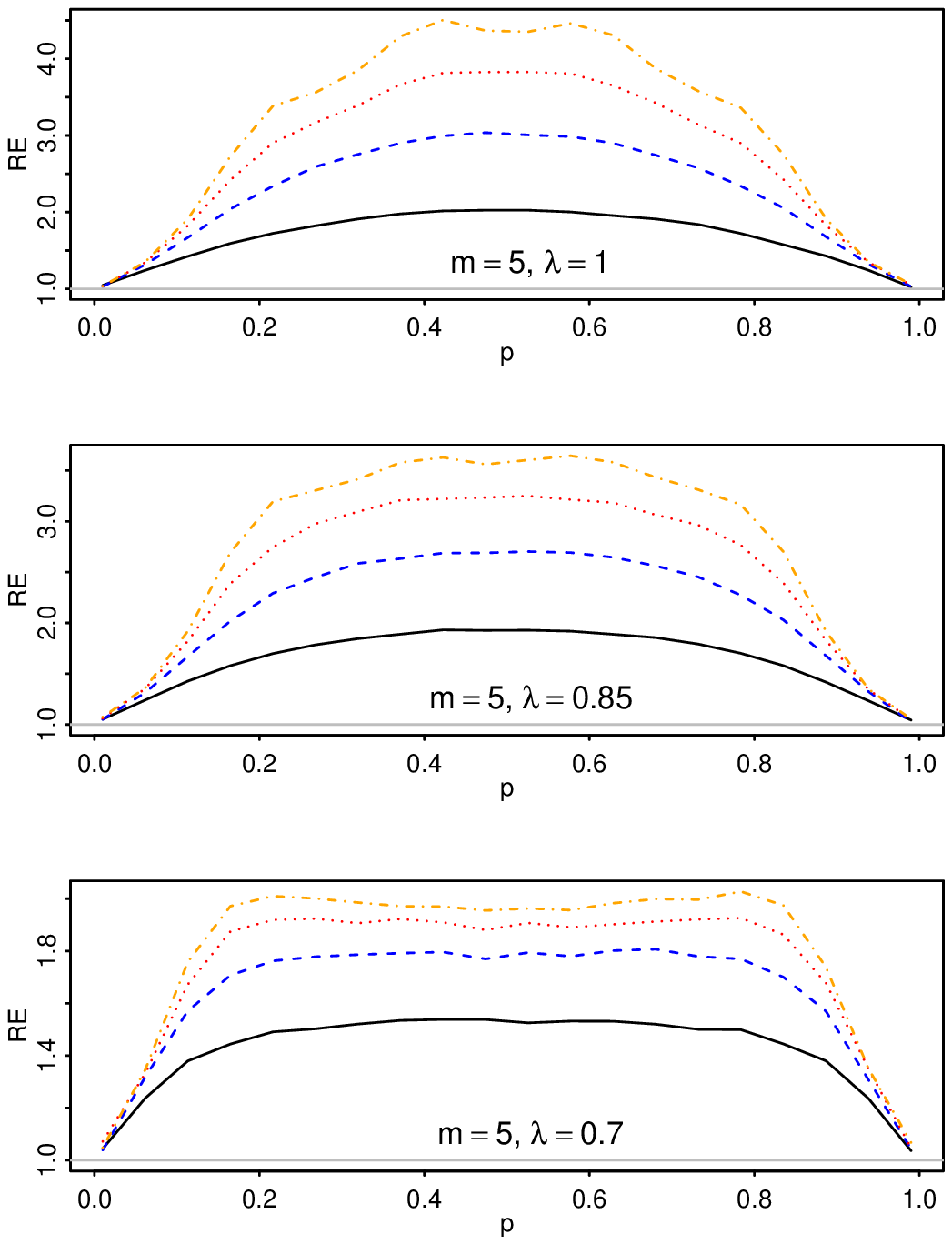}
\caption{The estimated RE for $m=5$ and $\lambda \in \{0.7,0.85,1\}$, where black/solid, blue/dashed, red/dotted, and orange/curves relate to $r=1$,  $r=2$, $r=3$, and $r=4$, respectively.}
\end{figure}

Suppose the variable of interest $X$, with mean $\mu_X$ and standard deviation $\sigma_X$, is ranked by means of a covariate $Y$. In this model, the two variables are related as
$$Y=\lambda \left( \frac{X-\mu_X}{\sigma_X} \right)+\sqrt{1-\lambda^2}\, Z,$$
where $Z$ is a standard normal random variable independent from $X$. Here, the ranking quality is determined by parameter $\lambda$, which is the correlation coefficient between $X$ and $Y$. The random ranking and the perfect ranking correspond to $\lambda=0$ and $\lambda=1$, respectively. In our problem, $X$ is a Bernoulli random variable with the success probability $p$, so we have $\mu_X=p$ and $\sigma_X=\sqrt{p(1-p)}$.

For each combinations of $m$, $r$, and $p$, efficiency of $\hat{p}_{\textrm{MSRSS}}^{(r)}$ relative to $\hat{p}_{\textrm{SRS}}$ was estimated, using the procedure described earlier. In doing so, the judgment ranking process was based on the above model with $\lambda \in \{0.7,0.85,1\}$. Figures 1-3 display the results for different set sizes. The proportion estimator in SRS is always outperformed by its MSRSS analog. Additionally, for fixed $m$ and $r$, the RE takes the highest value at $p=0.5$, while it declines symmetrically toward $p=0$ and $p=1$. For any combination of $m$ and $p$, increasing $r$ gives rise to improvement in the RE, and this is more evident as $p$ deviates from zero/unity. Finally, efficiency gain is naturally increasing in $\lambda$, given that the other factors are fixed. This is the case with most of the statistical methods developed for RSS-based schemes. It is interesting to note that for a small sample of size 5, $\hat{p}_{\textrm{MSRSS}}^{(r)}$ could be four times more efficient than $\hat{p}_{\textrm{SRS}}$ if $p$ is close to 0.5, and the perfect ranking is assumed (see the top panel in Figure 3).

As shown in Proposition 2 (b), the proportion estimator in MSRSS is more efficient than its SRS version based on the same number of measurements. Therefore, the proposed estimator has a good potential to be used whenever cost consideration is of high importance, which is the case in many medical studies. For example, if a predetermined error bound in estimating the true proportion is desired, then it can be achieved with a smaller sample in MSRSS as compared with SRS, thereby reducing the involved cost. Formally, percentage of sample size reduction (PSSR) can be measured via
$$\textrm{PSSR}=\left[1-\frac{\textrm{Var}\!\left( \hat{p}_{\textrm{MSRSS}}^{(r)} \right)}{\textrm{Var}\!\left( \hat{p}_{\textrm{SRS}} \right)} \right]\times 100.$$

Table 1 contains estimated values of this criterion for $m \in \{3,4,5\}$, $r \in \{1,2,3,4\}$, and $p \in \{0.1,0.25,0.5,0.75,0.9\}$ in the perfect ranking setup. The estimation is done similar to that of the RE. It is observed that trends of the RE in Figures 1-3 are reflected here. For fixed $m$ and $r$, the PSSR reaches its maximum at $p=0.5$, while it falls symmetrically toward $p=0.1$ and $p=0.9$. If the other factors are fixed, then the PSSR is increasing $r$. Interestingly, the proportion estimation in MSRSS needs 76.90\% fewer observations than that in SRS when $m=5$, $r=4$, and $p=0.5$. This could be an appealing feature for medical researchers whose works require expensive measurements.

\begin{table}
\centering
\caption{\small The estimated PSSR for $m \in \{3,4,5\}$, $r \in \{1,2,3,4\}$, and $p \in \{0.1,0.25,0.5,0.75,0.9\}$ in the perfect ranking setup.}
\begin{tabular}{c c c c c c c}\hline
& & \multicolumn{5}{c}{$p$}\\
\cline{3-7}
{$m$} & {$r$} & {0.1} & {0.25} & {0.5} & {0.75} & {0.9} \\ \hline
 3 & 1 & 16.16 & 30.88 & 37.37 & 30.78 & 16.61 \\
   & 2 & 20.52 & 43.87 & 53.20 & 43.25 & 20.47 \\
   & 3 & 21.53 & 50.73 & 59.78 & 50.90 & 21.71 \\
   & 4 & 22.04 & 55.17 & 63.19 & 55.41 & 22.04 \\
 & & & & & & \\
 4 & 1 & 22.73 & 38.53 & 45.70 & 38.55 & 22.38 \\
   & 2 & 28.89 & 53.59 & 61.21 & 53.33 & 29.22 \\
   & 3 & 31.57 & 61.80 & 69.34 & 61.49 & 31.40 \\
   & 4 & 32.71 & 67.24 & 73.96 & 67.34 & 32.52 \\
 & & & & & & \\
 5 & 1 & 27.94 & 44.51 & 50.72 & 44.26 & 27.80 \\
   & 2 & 36.84 & 60.38 & 66.99 & 60.16 & 37.03 \\
   & 3 & 40.61 & 67.24 & 73.68 & 67.47 & 40.96 \\
   & 4 & 42.53 & 71.32 & 76.90 & 71.51 & 42.73 \\
\hline
\end{tabular}
\end{table}

\subsection{Illustration using real data}
In this sub-section, the proposed procedure is illustrated using Wisconsin Breast Cancer Data (WBCD), which was originally compiled by Street et al. (1993). It is one of the first data sets where feature extraction was conducted in an attempt to apply a machine learning algorithm for improving malignancy prediction of a medical condition. Employing image recognition and machine learning techniques on this data set, Street et al. (1993) achieved a major advance in the accuracy of malignant tumor prediction.

The WBCD includes 699 observations on ten variables, which is accessible via ``mlbench" Package\footnote{https://cran.r-project.org/web/packages/mlbench/index.html} developed for R statistical software. The dichotomous variable of interest ($X$) indicates whether a tumor is malignant (success) or benign (failure). Here, malignancy was diagnosed through a comprehensive biopsy procedure. Additionally, there are nine visually assessed cytological covariates which are pertinent to determination of breast cancer: clump thickness ($Y_1$), uniformity of cell size ($Y_2$), uniformity of cell shape ($Y_3$), marginal adhesion ($Y_4$), single epithelial cell size ($Y_5$), bare nuclei ($Y_6$), bland chromatin ($Y_7$), normal nucleoli ($Y_8$), and mitoses ($Y_{10}$). These covariates are easily obtained from the FNA biopsy, and their values range from 1 (normal) through 10 (most abnormal).

In the following, the WBCD is considered as a hypothetical population. Then, drawing a multistage ranked set sample is exemplified, and efficiency of SRS and MSRSS in estimating the population proportion are compared.

\subsubsection{An example of sampling in MSRSS}
Here, we describe MSRSS using $r=2$, $m=3$, and $n=1$. It is assumed that the judgment ranking is based on the covariate $Y_2$. First, 27 sample units are drawn with replacement from the population, and randomly divide them into 3 sets of size 9. The three sets are given by
\begin{equation*}
\begin{pmatrix}
V_{11}^1 (8) & V_{12}^1 (1) & V_{13}^1 (1) \\
V_{21}^1 (4) & V_{22}^1 (2) & V_{23}^1 (1) \\
V_{31}^1 (1) & V_{32}^1 (6) & V_{33}^1 (8)
\end{pmatrix}, \quad
\begin{pmatrix}
V_{11}^2 (10) & V_{12}^2 (6) & V_{13}^2 (1) \\
V_{21}^2 (1) & V_{22}^2 (7) & V_{23}^2 (1) \\
V_{31}^2 (2) & V_{32}^2 (1) & V_{33}^2 (1)
\end{pmatrix}, \quad
\begin{pmatrix}
V_{11}^3 (1) & V_{12}^3 (8) & V_{13}^3 (1) \\
V_{21}^3 (3) & V_{22}^3 (6) & V_{23}^3 (1) \\
V_{31}^3 (1) & V_{32}^3 (6) & V_{33}^3 (3)
\end{pmatrix},
\end{equation*}
where we utilized the same notation introduced in Section 2 for illustration of MSRSS. Also, value of the covariate for any unit appears in parentheses. In each set, the units of each row are ordered using the covariate's information. In the case that ties occur during this process, they are broken at random. Proceeding in this way, we arrive at
\begin{equation*}
\begin{pmatrix}
\boxed{V_{[11]}^1 (1)} & V_{[12]}^1 (1) & V_{[13]}^1 (8) \\
V_{[21]}^1 (1) & \boxed{V_{[22]}^1 (2)} & V_{[23]}^1 (4) \\
V_{[31]}^1 (1) & V_{[32]}^1 (6) & \boxed{V_{[33]}^1 (8)}
\end{pmatrix}, \quad
\begin{pmatrix}
\boxed{V_{[11]}^2 (1)} & V_{[12]}^2 (6) & V_{[13]}^2 (10) \\
V_{[21]}^2 (1) & \boxed{V_{[22]}^2 (1)} & V_{[23]}^2 (7) \\
V_{[31]}^2 (1) & V_{[32]}^2 (1) & \boxed{V_{[33]}^2 (2)}
\end{pmatrix}, \quad
\begin{pmatrix}
\boxed{V_{[11]}^3 (1)} & V_{[12]}^3 (1) & V_{[13]}^3 (8) \\
V_{[21]}^3 (1) & \boxed{V_{[22]}^3 (3)} & V_{[23]}^3 (6) \\
V_{[31]}^3 (1) & V_{[32]}^3 (3) & \boxed{V_{[33]}^3 (6)}
\end{pmatrix}.
\end{equation*}

Now, the units in the sets
$$S_1=\left\{ V_{[11]}^1 (1), V_{[22]}^1 (2), V_{[33]}^1 (8) \right\}, \quad S_2=\left\{ V_{[11]}^2 (1), V_{[22]}^2 (1), V_{[33]}^2 (2) \right\}, \quad S_3=\left\{ V_{[11]}^3 (1), V_{[22]}^3 (3), V_{[33]}^3 (6) \right\},$$
are ordered, and the unit with judgment rank $i$ ($i=1,2,3$) in $S_i$ is selected for actual measurement. This is to say that the 2nd stage ranked set sample is obtained by quantifying $X$ for $V_{[11]}^1$, $V_{[22]}^2$, and $V_{[33]}^3$. The resulting sample is given by $\left\{X_{[1]1}^{(2)}=0,X_{[2]1}^{(2)}=0,X_{[3]1}^{(2)}=1\right\}$, where 0 (1) shows that a tumor is benign (malignant). According to MSRSS procedure in Section 2, $X_{[i]j}^{(r)}$ is the $i$th judgment order statistic in the $j$th cycle.

\subsubsection{Efficiency comparison}
In this sub-section, we investigate performances of SRS and MSRSS designs in estimating the proportion of malignant breast tumors. In this population, 241 out of 699 patients have malignant breast tumors, so the true population proportion is $p=0.34$. The efficiency comparison is based on the RE defined in Section 3.1. This quantity is again determined through Monte Carlo simulation with 100,000 replications. In generating multistage ranked set samples, sampling from the WBCD is performed with replacement to guarantee that the quantified units are independent of each other. Also, the judgment ranking can be based on any of the nine covariates described earlier. In particular, we utilized $Y_2$, $Y_5$, and $Y_9$. Spearman correlation coefficient for pairs $(X,Y_2)$, $(X,Y_5)$, and $(X,Y_9)$ are 0.86, 0.76, and 0.53, respectively. Therefore, these choices of the covariate allow us to evaluate effect of the ranking quality on the performance of $\hat{p}_{\textrm{MSRSS}}^{(r)}$.

Figure 4 shows the estimated RE for $m \in \{3,4,5\}$ and $r \in \{1,2,3,4\}$, under the above three judgment ranking scenarios. We used $n=1$ because the RE is independent of $n$, as mentioned before. It is seen that values of the RE exceed unity in all of the situations considered. Accordingly, the proportion estimator in MSRSS is more efficient than its competitor in SRS based on the same number of measurements. Moreover, for a fixed $m$, the RE is increasing in $r$. As expected, using $Y_2$ leads to larger values of the RE. It is worth noting that among the nine covariates, $Y_2$ has the highest correlation coefficient with the variable of interest $X$.

\begin{figure}[h!]
\centering
\includegraphics[]{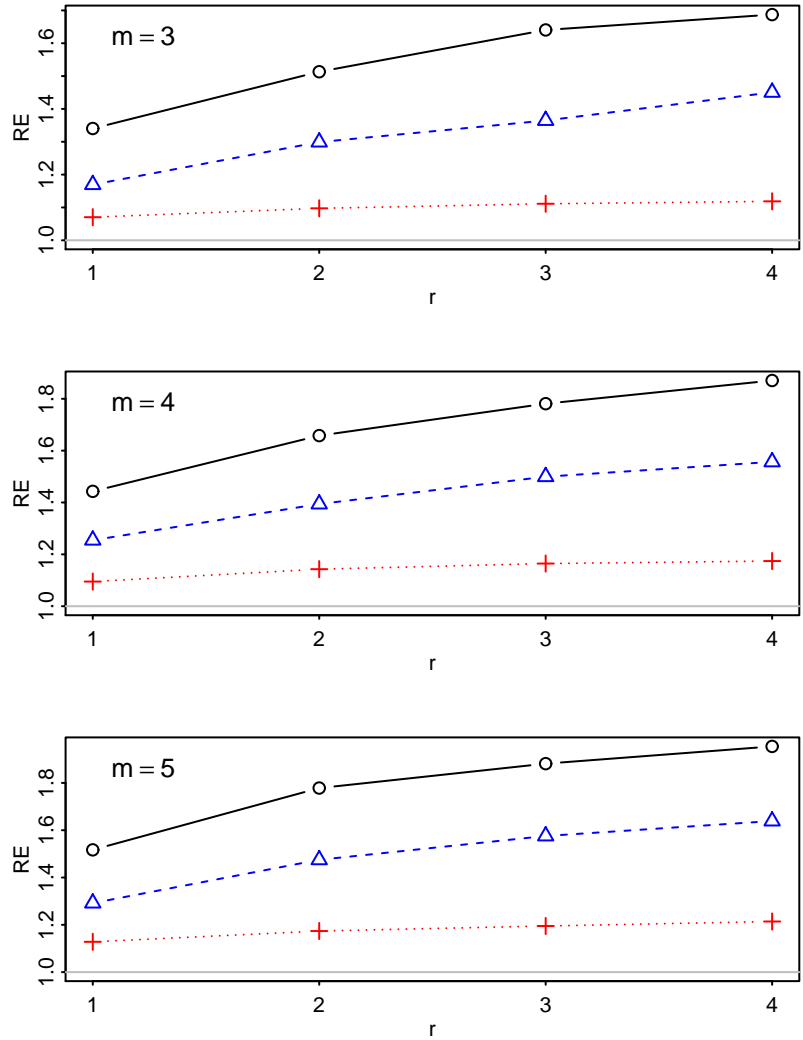}
\caption{The estimated RE from the WBCD for $m \in \{3,4,5\}$ and $r \in \{1,2,3,4\}$, where black/solid, blue/dashed, and red/dotted curves relate to covariates  $Y_2$, $Y_5$, and $Y_9$, respectively.}
\end{figure}

It is interesting to examine the PSSR in estimating the proportion from the WBCD. Table 2 reports estimated values of the PSSR for $m \in \{3,4,5\}$ and $r \in \{1,2,3,4\}$. The estimation is based on 100,000 samples, and $Y_2$ is used in the judgment ranking process. Given a set size, the PSSR improves as the stage number becomes larger. This property was also observed for each $p$ in Table 1. Finally, estimated values of the PSSR fall in the range 25\%-49\%, which seems fairly good.

\begin{table}
\centering
\caption{\small The estimated PSSR from the WBCD for $m \in \{3,4,5\}$ and $r \in \{1,2,3,4\}$, when covariate $Y_2$ is used.}
\begin{tabular}{c c c c c}\hline
& \multicolumn{4}{c}{$r$}\\
\cline{2-5}
{$m$} & {1} & {2} & {3} & {4} \\ \hline
 3 & 25.33 & 34.06 & 38.59 & 40.40 \\
 4 & 30.11 & 39.29 & 44.34 & 46.52 \\
 5 & 33.90 & 43.27 & 47.07 & 48.89 \\
\hline
\end{tabular}
\end{table}

\section{Discussion}
Our developed procedure is quite general and can be applied to estimate a proportion in other problems. For example, suppose we want to study prevalence of obesity in a population, based on body fat. Dual energy X-ray absorptiometry is one of the body fat testing methods that has been validated and thus, is considered as the ``gold standard", but it is too costly to implement. The abdomen circumference is an obesity measure which is obtained readily. Thus, it can be employed as a covariate for using our proportion estimator in studying obesity. We believe that similar situations are abundant in medicine.

A deficiency associated with MSRSS design is that drawing a sample of size $m$, in a single cycle, requires identifying $m^{r+1}$ units from the population. The number of units may be big if $m$ and $r$ are large. This might hinder the use of this design in practice. To overcome this problem, we plan to use multistage pair ranked set sampling in estimating the proportion. It is a modification of MSRSS, due Mahdizadeh and Zamanzade (2019b), that allows cutting down the number of units needed for preparatory rankings by almost half.
\section{Conclusion}
This article puts forward an efficient method for estimating prevalence of breast cancer. It builds on MSRSS that incorporates auxiliary information in order to guide the experimenter toward drawing a more representative sample, as compared with SRS. Theoretical properties of the proposed estimator are investigated, and some numerical studies are performed.

In some medical studies, it is of interest to make inference about a population's characteristic, when measurement is expensive and/or time-consuming. Utilizing cost-efficient statistical methods for data analysis is critical at this juncture. We hope that our proposal would be a powerful tool in the arsenal of medical researchers.


\section*{References}
\begin{description}

\item  Al-Saleh, M.F., and Al-Omari, A.I. (2002). Multistage ranked set sampling. J Stat Plan Inference 102, 273-286.

\item Bouza-Herrera, C.N., and Al-Omari, A.I. (2019). Ranked Set Sampling, 65 Years Improving the Accuracy in Data Gathering. Academic Press.

\item  Chen, H., Stasny, E.A., and Wolfe, D.A. (2005). Ranked set sampling for efficient estimation of a population proportion. Stat Med 24, 3319-3329.

\item  Dell, T.R., and Clutter, J.L. (1972). Ranked set sampling theory with order statistics background. Biometrics 28, 545-555.

\item  Frey, J. (2007). New imperfect rankings models for ranked set sampling. J Stat Plan Inference 137, 1433-1445.

\item  Frey, J., and Feeman, T.J. (2018). Finding the maximum efficiency for multistage ranked-set sampling. Commun Stat Theory Methods 47, 4131-4141.

\item  Gemayel, N.M., Stasny, E.A., Tackett, J.A., and Wolfe, D.A. (2012). Ranked set sampling: An auditing application. Review of Quantitative Finance and Accounting 39, 413-422.

\item  Mahdizadeh, M., and Zamanzade, E. (2017a). Estimation of a symmetric distribution function in multistage ranked set sampling. Statistical Papers 61, 851-867.

\item  Mahdizadeh, M., and Zamanzade, E. (2017b) Reliability estimation in multistage ranked set sampling. Revstat Stat J 15, 565-581.

\item  Mahdizadeh, M., and Zamanzade, E. (2018). Smooth estimation of a reliability function in ranked set sampling. Statistics 52, 750-768.

\item  Mahdizadeh, M., and Zamanzade, E. (2019a). Dynamic reliability estimation in a rank-based design. PROBAB MATH STAT-POL 39, 1-18.

\item  Mahdizadeh, M., and Zamanzade, E. (2019b). Efficient body fat estimation using multistage pair ranked set sampling. Stat Meth Med Res 28, 223-234.

\item  McIntyre, G.A. (1952). A method of unbiased selective sampling using ranked sets. Aust J Agric Res 3, 385-390.

\item  McIntyre, G.A. (2005). A method of unbiased selective sampling using ranked sets. Am Stat 59, 230-232.

\item  Presnell, B., and Bohn, L.L. (1999). U-Statistics and imperfect ranking in ranked set sampling. Journal J Nonparametr Stat 10, 111-126.

\item Shapiro, C.L. (2018). Cancer Survivorship. New England Journal of Medicine 379, 2438-2450.

\item  Street, W.N., Wolberg, W.H., and Mangasarian, O.L. (1993). Nuclear feature extraction for breast tumor diagnosis. Proceedings of SPIE - The International Society for Optical Engineering 1905, 861-870.

\item  Terpstra, J.T., and Liudahl, L.A. (2004). Concomitant-based rank set sampling proportion estimates. Stat Med 23, 2061-2070.

\item  Terpstra, J.T., and Miller, Z.A. (2006). Exact inference for a population proportion based on a ranked set sample. Commun Stat Simul Comput 35, 19-27.

\item  Terpstra, J.T., and Wang, P. (2008). Confidence intervals for a population proportion based on a ranked set sample. J Stat Comput Simul 78, 351-366.

\item  Wolfe, D.A. (2012). Ranked set sampling: Its relevance and impact on statistical inference. ISRN Probability and Statistics 1-32.

\item  Zamanzade, E., and Mahdizadeh, M. (2017). A more efficient proportion estimator in ranked set sampling. Stat Probab Lett 129, 28-33.

\end{description}

\section*{Appendix A}
{\bf Proof of Proposition 1.} a) Let $F_{[i]}(x)$ ($i=1,\ldots,m$) be the common distribution function of judgment order statistics with rank $i$, i.e. $X_{[i]1},\ldots,X_{[i]n}$. If the ranking scheme is consistent, then we have
\begin{equation}\label{iden}
F(x)=\frac{1}{m} \sum_{i=1}^m F_{[i]}(x),
\end{equation}
where $F(x)$ is the population distribution function (see Lemma 1 in Presnell and Bohn (1999) for details). The unbiasedness of $\hat{p}_{\textrm{RSS}}$ is immediate from this identity. Also, the variance expression is obtained by noting that elements of $\{X_{[i]j}: i=1,\ldots,m\,;j=1,\ldots,n \}$ are independent, and each $X_{[i]j}$ is a Bernoulli random variable with the success probability $p_{[i]}$.

\noindent b) From the first part, one can write
\begin{eqnarray*}
Var\left( \hat{p}_{\textrm{RSS}} \right)&=& \frac{1}{n m^2} \sum_{i=1}^m p_{[i]} \left( 1-p_{[i]} \right) \\
&=& \frac{1}{n m^2} \left[ \sum_{i=1}^m p_{[i]}-\sum_{i=1}^m p_{[i]}^2 \right] \\
&=& \frac{1}{n m^2} \left[ mp-\sum_{i=1}^m \left( p_{[i]}-p+p \right)^2 \right] \\
&=& \frac{1}{n m^2} \left[ mp-mp^2-\sum_{i=1}^m \left( p_{[i]}-p \right)^2 \right] \\
&=& \frac{p(1-p)}{n m}-\frac{1}{n m^2} \sum_{i=1}^m \left( p_{[i]}-p \right)^2 \\
&\leq& Var\left( \hat{p}_{\textrm{SRS}} \right),
\end{eqnarray*}
where in the third equality, identity \eqref{iden} has been used.

\noindent c) It can be seen that
$$\hat{p}_{\textrm{RSS}}=\frac{1}{n}\left( \frac{1}{m}\sum_{i=1}^m X_{[i]1} +\cdots+ \frac{1}{m}\sum_{i=1}^m X_{[i]n} \right).$$
To put it another way, $\hat{p}_{\textrm{RSS}}$ is the sample mean of $n$ independent and identically distributed random variables with mean $p$ and variance $\sum_{i=1}^m p_{[i]} \left( 1-p_{[i]} \right)/m^2$. The asymptotic normality is then concluded from the central limit theorem. \quad $\Box$

{\bf Proof of Proposition 2.} a) An identity similar to \eqref{iden} holds in MSRSS (see Proposition 3 in Mahdizadeh and Zamanzade (2017a)). It establishes that
\begin{equation}\label{iden2}
F(x)=\frac{1}{m} \sum_{i=1}^m F_{[i]}^{(r)}(x),
\end{equation}
where $F_{[i]}^{(r)}(x)$ ($i=1,\ldots,m$) is the common distribution function of $X_{[i]1}^{(r)},\ldots,X_{[i]n}^{(r)}$. An application of identity \eqref{iden2} shows the unbiasedness of $\hat{p}_{\textrm{MSRSS}}^{(r)}$. The corresponding variance is derived similar to that of $\hat{p}_{\textrm{RSS}}$.

\noindent b) This is proved in line with part (b) of Proposition 1.

\noindent c) Suppose $\mathcal{X}_{(i)}^{(r-1)}$ ($i=1,\ldots,m$) is the $i$th order statistic of $X_{[1]1}^{(r-1)},\ldots,X_{[m]1}^{(r-1)}$. Then, we have
\begin{eqnarray}\label{p1}
Var\left( \hat{p}_{\textrm{MSRSS}}^{(r-1)} \right)&=&\frac{1}{n m^2} Var\left( \sum_{i=1}^m X_{[i]1}^{(r-1)} \right) \nonumber \\
&=& \frac{1}{n m^2} Var\left( \sum_{i=1}^m \mathcal{X}_{(i)}^{(r-1)} \right) \nonumber \\
&=& \frac{1}{n m^2} \left[ \sum_{i=1}^m Var\left( \mathcal{X}_{(i)}^{(r-1)} \right)+ \sum_{i\neq j=1}^m  Cov\left( \mathcal{X}_{(i)}^{(r-1)}, \mathcal{X}_{(j)}^{(r-1)} \right)\right].
\end{eqnarray}
It can be shown that the covariance terms in \eqref{p1} are positive. To do so, without loss of generality, it is assumed that $i<j$. Then, one can write
\begin{eqnarray}\label{p2}
Cov\left( \mathcal{X}_{(i)}^{(r-1)}, \mathcal{X}_{(j)}^{(r-1)} \right)&=& E\left( \mathcal{X}_{(i)}^{(r-1)} \mathcal{X}_{(j)}^{(r-1)} \right)-E\left( \mathcal{X}_{(i)}^{(r-1)} \right) E\left( \mathcal{X}_{(j)}^{(r-1)} \right) \nonumber \\
&=& P\left( \mathcal{X}_{(i)}^{(r-1)}=1 \right)-P\left( \mathcal{X}_{(i)}^{(r-1)}=1 \right) P\left( \mathcal{X}_{(j)}^{(r-1)}=1 \right) \nonumber \\
&=& p_{[i]}^{(r)} \left( 1-p_{[j]}^{(r)} \right).
\end{eqnarray}
Putting \eqref{p1} and \eqref{p2} together, it follows that
\begin{eqnarray*}
Var\left( \hat{p}_{\textrm{MSRSS}}^{(r-1)} \right)&\geq & \frac{1}{n m^2} \sum_{i=1}^m Var\left( \mathcal{X}_{(i)}^{(r-1)} \right) \nonumber \\
&=& \frac{1}{n m^2} \sum_{i=1}^m Var\left( X_{[i]1}^{(r)} \right) \nonumber \\
&=& Var\left( \hat{p}_{\textrm{MSRSS}}^{(r)} \right),
\end{eqnarray*}
where the first equality results from the fact that $\mathcal{X}_{(i)}^{(r-1)}$ and $X_{[i]1}^{(r)}$ are identically distributed according to MSRSS procedure.

\noindent d) Proof of asymptotic normality of $\hat{p}_{\textrm{MSRSS}}^{(r)}$ parallels that of $\hat{p}_{\textrm{RSS}}$, and it is omitted. \quad $\Box$

\end{document}